\documentclass[fleqn,10pt]{SelfArx} 

\usepackage[english]{babel} 
\usepackage{natbib}
\usepackage{multicol}
\usepackage{pdfpages}
\titlespacing\section{0pt}{4pt plus 4pt minus 2pt}{0pt plus 2pt minus 2pt}
\titlespacing\subsection{0pt}{4pt plus 4pt minus 2pt}{0pt plus 2pt minus 2pt}
\titlespacing\subsubsection{0pt}{4pt plus 4pt minus 2pt}{0pt plus 2pt minus 2pt}

\newcommand{\apjl}{ApJL}

\newcommand{\ao}{Appl. Optics}
\newcommand{\aap}{A\&A}

\definecolor{color1}{RGB}{0,0,90} 
\definecolor{color2}{RGB}{0,20,20} 

\usepackage{hyperref} 
\hypersetup{hidelinks,colorlinks,breaklinks=true,urlcolor=color2,citecolor=color1,linkcolor=color1,bookmarksopen=false,pdftitle={Title},pdfauthor={Author}}

\JournalInfo{NAS Exoplanet Whitepaper, 2018} 
\Archive{} 

\PaperTitle{Direct Imaging in Reflected Light: Characterization of Older, Temperate Exoplanets With 30-m Telescopes} 

\Authors{\'Etienne Artigau, Bjorn Benneke, Laird Close, Ian Crossfield, Jacques-Robert Delorme, Michael Fitzgerald\textsuperscript{1}, Jonathan Fortney, Andrew Howard, \'Etienne Artigau, Richard Frazin\textsuperscript{4}, Nemanja Jovanovic, Julien Lozi, Jared Males\textsuperscript{2}, Christian Marois, Benjamin A. Mazin\textsuperscript{3}, Max A. Millar-Blanchaer, Gautam Vasisht, J. Kent Wallace, Ji Wang} 
\affiliation{\textsuperscript{1}\textit{Department of Physics and Astronomy, University of California, Los Angeles, CA}} 
\affiliation{\textsuperscript{2}\textit{Department of Astronomy and Steward Observatory, University of Arizona}}
\affiliation{\textsuperscript{3}\textit{Department of Physics, University of California, Santa Barbara, CA}}
\affiliation{\textsuperscript{4}\textit{Department of Climate and Space Sciences and Engineering, University of Michigan, Ann Arbor, MI}}

\affiliation{*\textbf{Corresponding author}: john@smith.com} 

\Keywords{Exoplanets --- Direct Imaging --- TMT --- GMT -- ELT} 
\newcommand{\keywordname}{Keywords} 

\begin{document}
\includepdf[pages={1}]{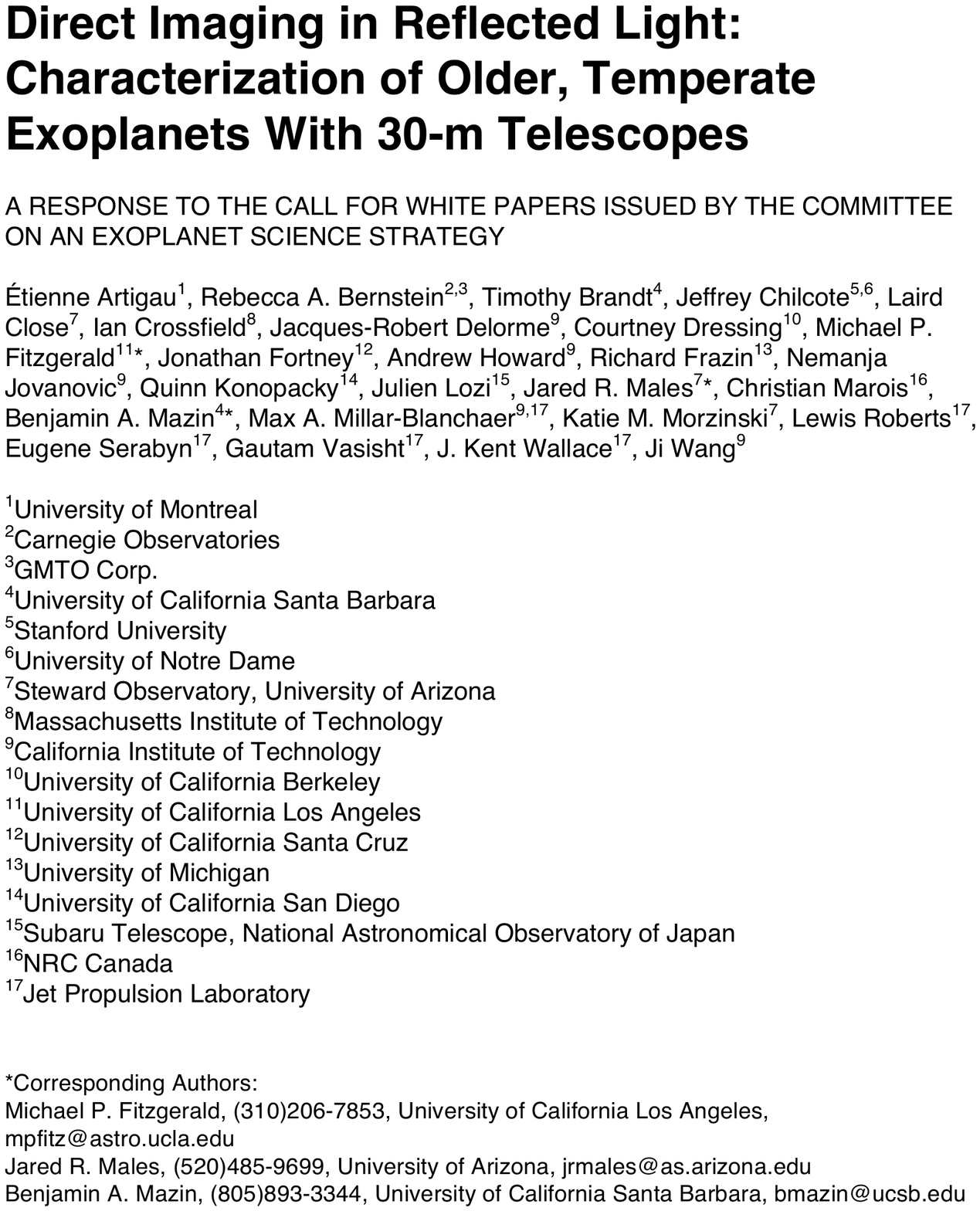}

\flushbottom 





\section*{Executive Summary}
Over the past three decades instruments on the ground and in space have discovered thousands of planets orbiting nearby stars. These observations have given rise to an astonishingly detailed picture of the demographics of short-period planets (P $<$ 10 days), but are incomplete at longer periods where both the sensitivity of transit surveys and radial velocity signals plummet. Even more glaring is that the spectra of planets discovered with these indirect methods are often inaccessible (most RV and all microlensing detections) or only available for a small subclass of transiting planets. Direct detection, also known as direct imaging, is a method for discovering and characterizing the atmospheres of planets at intermediate and wide separations.  It is the only means of obtaining spectra of non-transiting exoplanets. Today, only a handful of exoplanets have been directly imaged, and these represent a rare class of young, self-luminous super-Jupiters orbiting tens to hundreds of AU from their host stars. Characterizing the atmospheres of planets in the $<5$ AU regime, where RV surveys have revealed an abundance of other worlds, requires a 30-m-class aperture in combination with an advanced adaptive optics system, coronagraph, and suite of spectrometers and imagers -- this concept underlies planned instruments for both TMT (the Planetary Systems Imager, or PSI) and the GMT (GMagAO-X). These instruments could provide astrometry, photometry, and spectroscopy of an unprecedented sample of rocky planets, ice giants, and gas giants. For the first time habitable zone exoplanets will become accessible to direct imaging, and these instruments have the potential to detect and characterize the innermost regions of nearby M-dwarf planetary systems in reflected light. High-resolution spectroscopy will not only illuminate the physics and chemistry of exo-atmospheres, but may also probe rocky, temperate worlds for signs of life in the form of atmospheric biomarkers (combinations of water, oxygen and other molecular species). By completing the census of non-transiting worlds at a range of separations from their host stars, these instruments will provide the final pieces to the puzzle of planetary demographics.  This whitepaper explores the science goals of direct imaging on 30-m telescopes and the technology development needed to achieve them.

\setcounter{page}{1}

\begin{multicols}{2}

\section{Introduction}\label{sec:intro}

Exoplanet direct imaging is necessarily the future of exoplanet science.  RV surveys have a very important role to play in discovering exoplanets and measuring their mass, but as an indirect detection method its science return has limited scope.  Transit surveys help to fill in this information, but are hampered by the small fraction of systems that transit and practical limits on atmospheric spectroscopy for atmospheres with small scale heights, especially for planets with long periods.

Direct sensing at high contrast allows us to directly image and obtain spectra of planets.  Despite the technique's obvious attractions, it is technically demanding.  The first generation of dedicated facility planet imagers, GPI and SPHERE, were optimistic in their assumptions about the achieved contrast and exoplanet populations at wide separations, and as with many first attempts encountered unforeseen issues.  Despite this, SPHERE has met its design contrast goals, and GPI should meet or exceed them with investment.  New instrumentation for 30-m telescopes (GSMTs, Giant Segmented Mirror Telescopes) will benefit from 10--20 years of progress in our understanding of direct imaging, significantly improved technology, a much clearer picture of the underlying demographics of exoplanets, and a decade of experimentation on testbeds like SCExAO and MagAO-X to improve contrast.

Direct imaging is beginning to climb the power law of the exoplanet distribution.  Given funding and the space to innovate, scientific progress will be rapid.

\section{Direct Characterization of Known Systems with GSMTs}\label{sec:science}

\begin{figure*}[!ht]\centering 
\vspace{-.3in}
\includegraphics[width=0.45\textwidth]{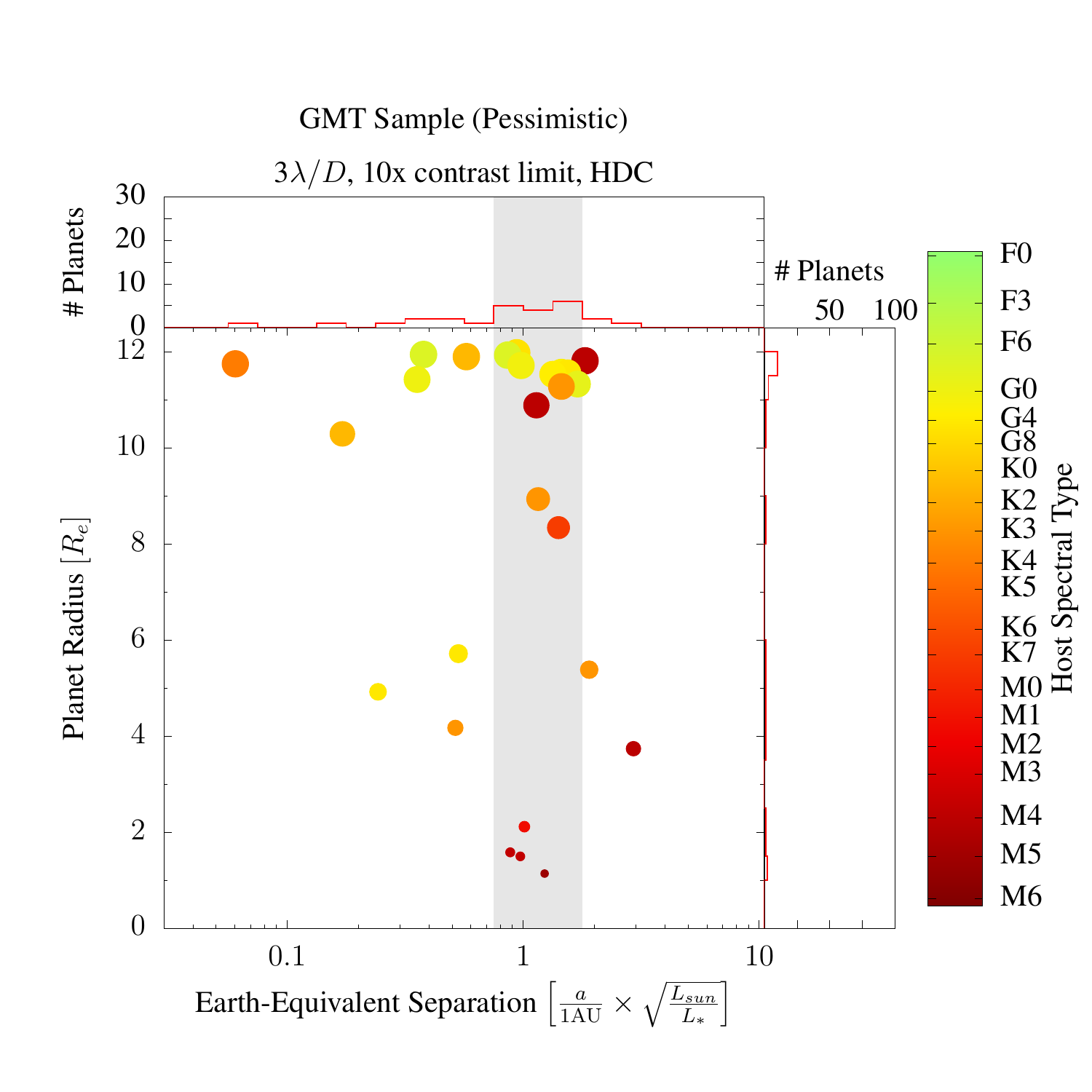}
\includegraphics[width=0.45\textwidth]{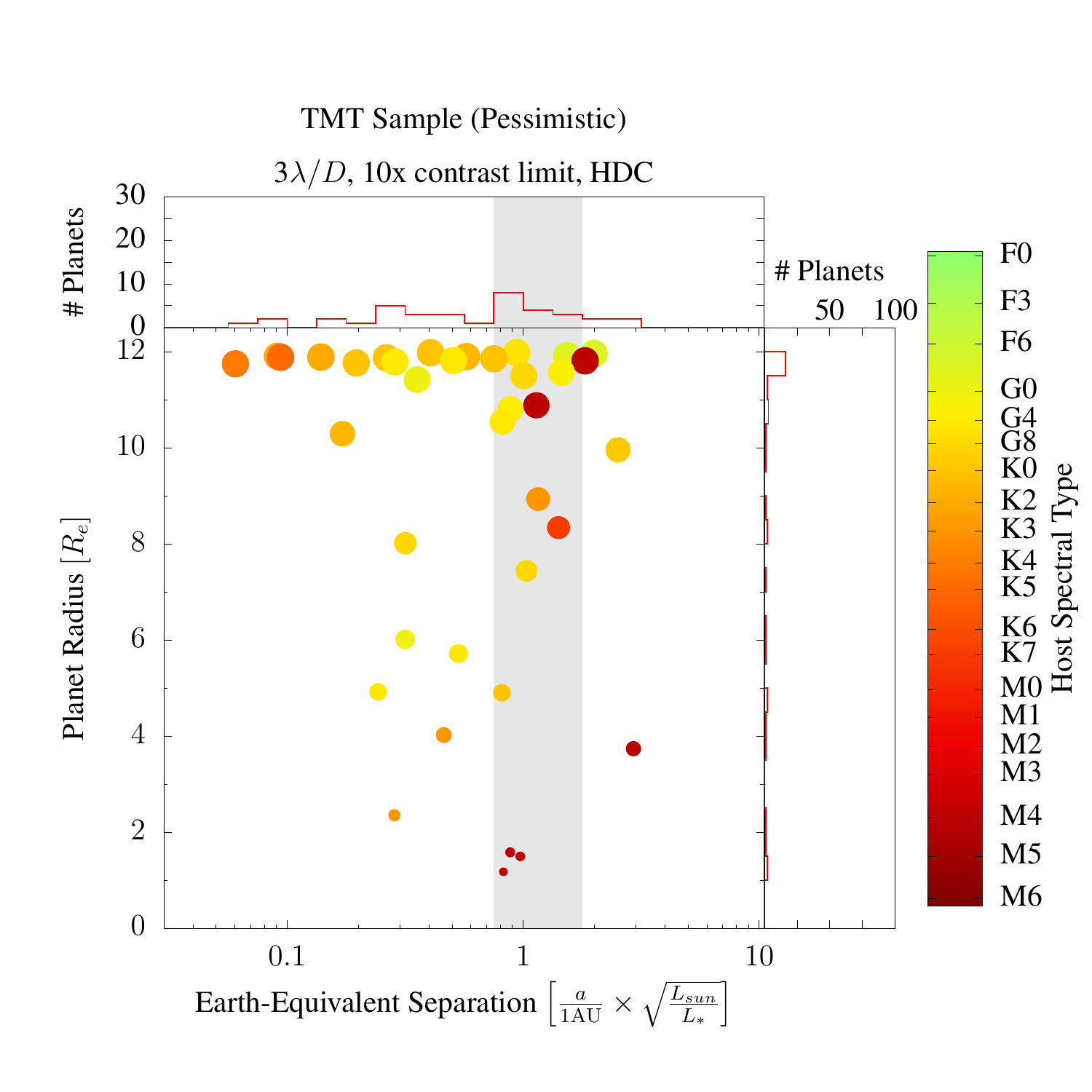}
\includegraphics[width=0.45\textwidth]{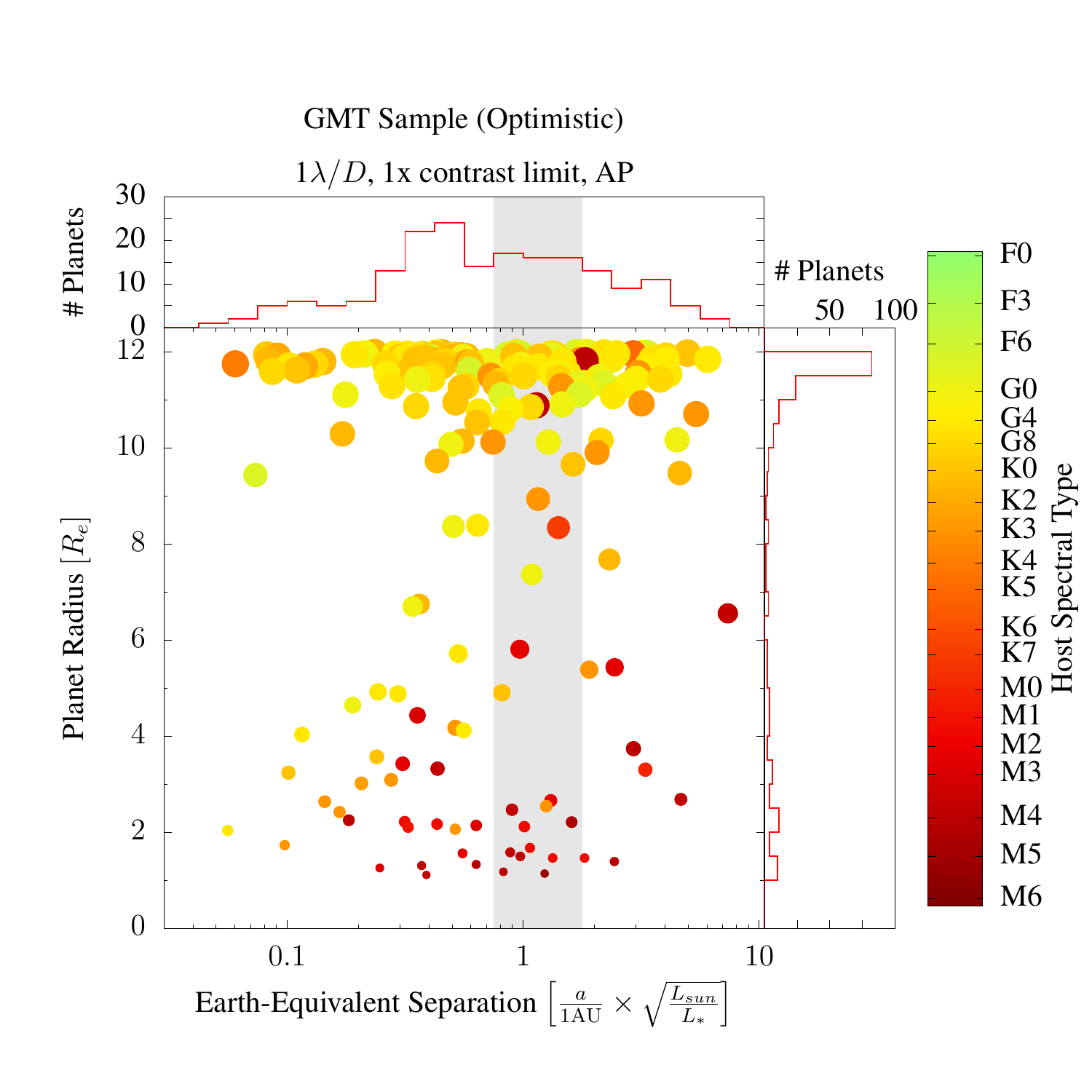}
\includegraphics[width=0.45\textwidth]{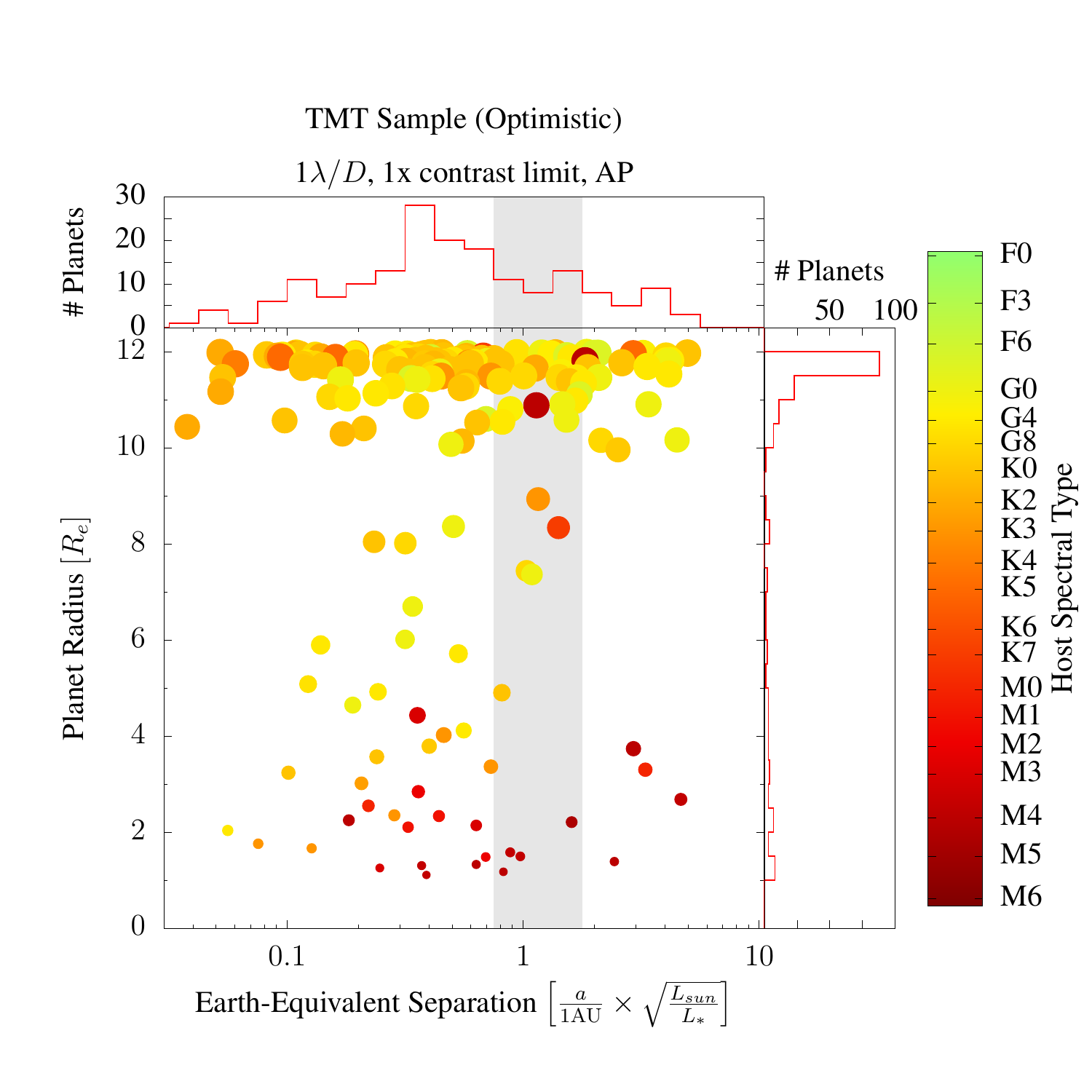}
\caption{The detectability of currently known planets (primarily detected through radial velocities) with GMT (left panel) and TMT (right panel) in a survey spanning 28 nights of integration.  The top panels assume the pessimistic case of a $3\lambda/D$ IWA coronagraph, and a residual stellar halo of 10$\times$ the limits of \cite{males_guyon_2018}. The residual speckles are assumed to be long lived, so the high-dispersion coronagraphy (HDC) technique is used to cross-correlate the reflected stellar spectrum.  Spectral-type specific templates were used to derive the cross-correlation signal boost.  Even with degraded instrument performance, several dozen known planets will be characterized in reflected light.  The bottom panels assume a 1 $\lambda/D$ inner working angle (IWA), photon noise, and residual short-lived speckle noise after on-line control of quasi-static aberrations and predictive control of the atmosphere.  Now aperture photometry (AP) is more efficient~(\S\ref{subsubsec:noise}).  An empirical planetary mass-radius relationship is used to derive planet radii.  Significant samples of both giant (R$_{\textrm{jup}}$=11 R$_{\textrm{earth}}$) and rocky planets are detected across a range of equilibrium temperatures, probing regions where condensates like H$_2$O, NH$_3$, and CH$_4$ are expected to play a major role in regulating planet formation. \textbf{These plots bound the potential of the coming GSMTs for reflected-light characterization of exoplanets, and should motivate significant efforts toward optimizing ground-based instruments for direct imaging.}}
\label{fig:jared}
\vspace{-.2in}
\end{figure*}



Of the thousands of exoplanets discovered, only a handful have been directly imaged. These represent a rare class of young, self-luminous super-Jupiters orbiting tens to hundreds of AU from their host stars \citep{bowler_pasp_2016}.  Many of the thousands of exoplanets we have discovered (primarily through RV surveys) are at contrast levels accessible even with today's instrumentation, but the inner working angle (IWA) imposed by the telescope's diffraction limit puts them too close to the star to see.  With contrast levels similar to that delivered by the first generation of dedicated instruments designed a decade ago (GPI and SPHERE), a 30-m aperture would yield significant science, as shown in the top panels of Figure~\ref{fig:jared}.  As we argue here, performance can be improved further.  With such improved performance, illustrated in the bottom panels, between GMT and TMT there are already over 300 targets. These are primarily gas giants but range in radius down to Earth radius, which have already been confirmed and await spectroscopic characterization with dedicated instrumentation on these GSMTs.  This is, however, the tip of the iceberg.  By the time GSMTs are commissioned, future RV surveys, \textit{TESS}, and \textit{GAIA} will have dramatically expanded the target list.

Moderate or high-resolution spectroscopy of these planets can probe the depths of multiple water and methane features, allowing models to recover carbon or oxygen abundance and in turn enabling integrated studies of these abundances vs.~planetary location, planetary mass, and stellar properties. Around near targets, GSMT’s sensitivity will reach down to sub-Neptune sized giant planets that will be discovered by upcoming Doppler and astrometric surveys. With the recent deprioritization of coronagraphic science on the \textit{WFIRST} mission, GSMTs with adaptive optics are the only technique likely to characterize mature giant planets at AU-scale separations into at least 2030 or beyond. 


\section{Potential Impact in the 2030s}\label{sec:futureimpact}

Over the next decade, the spectroscopic exploration of transiting planets will advance rapidly with the combination of \textit{TESS} and \textit{JWST}, including $>$100 giant planets (albeit biased towards planets with high effective temperatures and/or planets orbiting low-mass stars). GSMTs, however, will directly image cooler planets and planets orbiting earlier-type stars.  This will probe different regimes of atmospheric chemistry and is subject to fewer model-dependent biases. Together, transit and direct imaging missions will provide the spectroscopy of the diverse array of planetary targets that are needed to complete our understanding of planet formation.

Looking past the exoplanets that are already known, GSMTs should be able to detect starlight reflected by rocky, habitable-zone exoplanets. These observations will be among the first opportunities to detect biosignatures in the atmospheres of other worlds.  While 10~$\mu$m observations may allow detection of rocky planets around a very select sample of the closest solar-type stars, planets around faint M-type stars are the most favorable targets for spectroscopic follow up with both GSMTs and \textit{JWST} (in the case of transiting planets). For example, a simultaneous detection of oxygen (O$_2$) and methane (CH$_4$) would be highly indicative of life. A detection of CH$_4$, however, is out of reach for \textit{JWST} given the low concentration of CH$_4$ and the relatively high mean molecular weight / small scale height of an Earth-like atmosphere. 


Planet Imagers on GSMTs will provide astrometry, photometry, and spectroscopy of an unprecedented sample of rocky planets, ice giants, and gas giants. For the first time habitable zone exoplanets will become accessible to direct imaging, as GSMTs have the potential to detect and characterize the innermost regions of nearby M-dwarf planetary systems in reflected light. These instruments' high-resolution spectroscopic capabilities will not only illuminate the physics and chemistry of exo-atmospheres, but may also probe rocky, temperate worlds for signs of life in the form of atmospheric biomarkers (combination of water, oxygen and other molecular species). By completing the census of non-transiting worlds at a range of separations from their host stars, they will also provide the final missing pieces to the puzzle of planetary demographics.

\section{Closing the Technology Gap at the Diffraction Limit}\label{sec:tech}
The challenge in reaching the full potential of the GSMTs for reflected-light spectroscopy appears two-fold, namely pushing contrast to more extreme values, and extending the effective IWA toward the diffraction limit of the telescope.  Fundamentally these stem from a single driving requirement, to adequately sense and control the lowest order modes of wavefront aberration.  Here we review major limitations of achieving contrast, with comparison  to existing systems, and discuss the ongoing technological developments that seek to increase contrast at the smallest angular separations.

\subsection{Contrast limits at small angular separations}
The adaptive optics wavefront control system and the coronagraphic starlight suppression system define the ``raw'' contrast level through the resulting the time-variable residual speckle field in the image plane.  However this is not the the contrast ratio that matters for detection and characterization.  Rather, the ``final'' contrast delivered by an instrument is the end result of the further suppression of starlight by the combination of back-end instrumentation and post-processing algorithms.  Progress can be made in improving achievable final contrast by attacking the wavefront correction fidelity and through post-coronagraph processing.

Contrast is a function of position relative to the host star, and to increase the contrast at a specific location (i.e. close to the star) we must first consider the image formation process.  Image formation in the high-contrast regime is more straightforward than in normal imaging through partially compensated turbulence; in the limit of low wavefront errors and coronagraphic suppression of light that is spatially coherent in the pupil, the long-exposure intensity at a given location is proportional to the power of the phase and amplitude errors at the corresponding spatial frequency in the pupil~\citep{sivaramakrishnan_etal02, perrin_etal03, macintosh_etal05, guyon05}.
Therefore in order to maximally suppress starlight near the diffraction limit of the telescope (at the smallest angular separations) we must accurately control the lowest-order spatial frequencies of aberration in the pupil.

\subsection{Wavefront Estimation Error}

In order for the AO system to properly compensate for the atmospheric distortion and telescope aberrations, it must first estimate the phase of the incoming wavefront from the wavefront sensor (WFS) telemetry.  In addition to the effects below, chromatic errors play a role, but are mitigated by sensing in the science band.

\subsubsection{Servo lag}
A major limitation to the current generation of dedicated high-contrast imaging instruments at small separations arises from servo lag.
The lag between the time the wavefront is sensed and the time the correction is applied results in residual wavefront error.  This is particularly pernicious for aberrations that evolve quickly, such as arising from turbulence.  For example, for frozen flow at velocity $\vec{v}$, the pupil-plane residual after a lag $\tau$ is given by $\Delta=\phi(\vec{x})-\phi(\vec{x}-\vec{v}\tau)$.  The residual power is $P_\Delta(\vec{k})=P_\phi(\vec{k})\sin^2(2\pi\tau\vec{v}\cdot\vec{k})$, and therefore has a scaling similar to the input power spectrum (e.g. von Karman) in which most power is at the lowest spatial frequencies $k$.  These speckles fluctuate on timescales comparable to those of the input atmospheric aberrations, and therefore average relatively quickly.  However, these speckles' Poisson fluctuations present a barrier to reaching the desired final contrasts.
At moderate wind speeds, the raw contrast limitations of GPI and SPHERE are due to servo-lag error at separations of many $\lambda/D$. 

\subsubsection{Non-common path aberrations (NCPAs)}


Perhaps equally critical are errors that are improperly sensed by the adaptive optics feedback loop, so-called non-common path aberrations (NCPAs).  These arise from instrumental aberrations, and can temporally vary on a range of timescales.  As a function of increasing variation timescales, optomechanical vibrations, drifts in the correct WFS reference points coupled with variations in sensor gain, and thermal drifts in optomechanics can all play a role in determining the degree of improperly sensed spatially coherent wavefront aberrations that result in NCPAs and degrade the raw contrast.  Whatever their causes, the NCPAs end up creating speckles in the image that can be difficult to distinguish from planets, and, the longer the timescale for evolution of these NCPA-induced speckles (quasi-static), the greater challenge in averaging down their variation.  




\subsection{Improving contrast at the smallest separations}
The current generation of facility class instruments, GPI and SPHERE, were built with the maximization of contrast at small IWAs in mind.  In the case of GPI, the calibration interferometer designed to coherently detect aberrations not common between the science path and its Shack-Hartmann WFS did not meet performance goals. 

We note that the pyramid WFS (PyWFS) is now generally considered to be the most promising WFS architecture for GSMTs due to its high sensitivity.  In terms of performance for a given guide-star magnitude, the PyWFS is much more sensitive than the Shack-Hartmann (used in GPI in SPHERE) for the low-order modes that are of primary importance for our case~\citep[for high-order modes, it is equally sensitive;][]{Ragazzoni_sensitivity1999, Esposito_Riccardi_AA2001, Chew_PySH_comparison_OptComm06, Viotto_PyramidBehavior2013}.

On-sky nulling of quasi-static speckles has been demonstrated with SPHERE and SCExAO ~\citep{fusco_etal15, martinache_etal14}.  However, both GPI and SPHERE were designed more than a decade ago; servo lag provides a practical limit to the closed-loop temporal bandwidth of their adaptive optics correction, and calibration of quasi-static errors in the science focal plane is still limited to low-bandwidth operation.

In the meantime, technical advances to addressing these problems have occurred in the laboratory and in on-sky instrumentation.  These approaches hold promise in increasing both the raw and final contrasts achievable with modern technology in the near term.  When coupled with the larger aperture of a GSMT and its improvements in contrast (i.e. the turbulent power per $\lambda/D$ element scales as $1/D^2$) and IWA, the science aims of~\S\S\ref{sec:science} \&~\ref{sec:futureimpact} are in reach.

\subsubsection{Control of rapidly varying wavefront errors}
Minimizing the speckles due to the atmosphere at small separations from the star requires that we address servo lag.  This can be partly mitigated by running the adaptive optics loop at higher frame rates, of order serveral kHz, an approach now feasible due to advances in computation and detector noise performance in the past decade.  However servo lag may not be limited by frame rate alone, as other loop delays can become significant.  Moreover, for a given stellar brightness, WFS noise increases with higher frame rate.

It has long been recognized that a more favorable approach is to instead apply a prediction of the future aberration to the deformable mirror~\citep{dessenne_etal98, dessenne_etal99, poyneer_etal07, johnson_etal11}.  At the penalty of increased computational needs, the previous wavefront measurements can be used to more accurately predict the needed correction.  A recent study focused on GSMTs~\citep{males_guyon_2018} shows the potential for large contrast gains on bright stars under frozen-flow turbulence.  Ultimately the effectiveness of the technique will depend on how quickly turbulence evolves away from linear prediction compared to the timescale needed to adequately sense the wavefront.

Another area of promise for improving our ability to correct rapidly variable instrumental wavefront errors, such as telescope vibrations, comes from the additional sensing of the optomechanical structures.  So-called ``sensor fusion'' techniques seek to incorporate additional data not derived from the starlight to predict wavefront error.  The additional sensor data can be incorporated into predictive control schemes, such as multi-rate Kalman filters~\citep{riggs16}.  \citet{guyon&males17} develop a scheme for incorporating both wavefront sensing and telescope accelerometer data to more accurately predict wavefront errors.

On-sky demonstrations of predictive control and sensor fusion on high-contrast instruments are needed to validate these techniques.  

\subsubsection{Sensing and control of quasi-static aberrations}

New schemes have arisen for the sensing of low-order quasi-static NCPAs affecting speckles at small separations. 
One class uses starlight that is normally rejected by the coronagraph to sense low-order mode aberrations; both the CLOWFS~\citep{guyon_etal09} and LLOWFS~\citep{singh_etal14, singh_etal15} architectures use starlight rejected in the coronagraph focal and Lyot planes to reduce, but not eliminate, low-order NCPA at higher bandwidths than allowed by traditional science cameras.  However true NCPA elimination must occur via the science focal plane.

The on-sky arrival of new low-noise fast-readout detectors, such as MKIDs and IR-APD arrays, have enabled much more powerful focal-plane wavefront sensing techniques.
\citet{Frazin-ShortExposureSelfConsistent} combines millisecond telemetry from the science camera and the WFS to estimate the NCPAs based on the modulation of speckles provided by rapidly varying residual atmospheric aberrations.  Similarly,~\citet{Codona-ShortExposure} propose to use this ms-timescale dual telemetry stream to sense the speckles arising from the NCPAs directly.

An evolutionary step in the real-time control of low-order NCPAs is the application of more active methods to probe aberration phase.  Ongoing efforts at SCExAO and Palomar using IR-APD and MKID cameras seek to use deformable mirror probe signals to estimate the amplitude and phase of speckles to facilitate coherent differential imaging and enable real-time speckle control.

\subsubsection{Increasing final contrast through backend instrumentation and post-processing}
In principle, techniques using short-exposure imaging from the science focal plane to determine the coherent phase of speckles (so they can be suppressed) can also be used for post-processing, albeit at increased computational cost.  Essentially these methods post-process the entire data stream, jointly estimating residual NCPAs and light that is incoherent with the star, effectively separating the quasi-static speckles from planet light.

Schemes relying on short-exposure images ultimately exploit the relative incoherence of starlight with planet's light.  Instead of relying on ms-timescale exposures, the self-coherent camera of~\citet{Delorme-SelfCoherentCamera} uses a hole in the coronagraph's Lyot stop to create fringes within the stellar speckles.  Being incoherent with the star, any planetary signals are unaltered and can be recovered through processing of the science image.

Another technique to separate light from a planet with that from the star is based on using a high-resolution spectrograph behind the coronagraph to cross-correlate (CC) a template with the planet's spectrum, in this case composed primarily of the reflected stellar spectrum~\citep{snellen_2015}.  This technique was recently dubbed High Dispersion Coronagraphy~\citep[HDC;][]{wang_HDCI_2017, mawet_HDCII_2017}, and has been demonstrated on-sky.  Several ongoing efforts now seek to link adaptive optics enabled coronagraphs with high-resolution spectrometer backends.

\subsubsection{Noise Sources and Observational Techniques}\label{subsubsec:noise}
The instrument architectures ultimately selected depend on our understanding of how to maximize final contrast.  Here we consider two techniques for recording images of exoplanets.  The first is standard aperture photometry (AP), conducted over a bandpass of width $W_\lambda$.  In this regime there are two sources of noise to consider: photon noise and speckle noise with statistical lifetime $\tau_{sl}$.  The second technique is HDC, and is not affected by speckle noise (in principle), but depends on the number of lines in the detected spectrum.

An analysis of these two techniques leads to the following expression for the ratio of the $S/N$ (Males et al., in prep)
\begin{equation}
\frac{S/N_\text{HDC}}{S/N_\text{AP}} = \frac{ \sqrt{\eta_{sp} }} {  \left[1-\frac{\Delta\lambda}{W_\lambda}\sum^{N_l} q_l \right]} \times \sqrt{\frac{\Delta\lambda N_{l,\text{eff}}}{W_\lambda}} \times \sqrt{1+C_H  F_* W_\lambda \tau_{sl}}
\label{eqn:hdc_ap}
\end{equation}
where $\eta_{sp}$ is the relative efficiency of the spectrograph, $\Delta \lambda$ is the resolution, $C_H$ is the residual contrast, and $F_*$ is the stellar flux at the focal plane. $N_l$ is the number of lines in the spectrum, with variable depths of $q_l$.  $N_{l,\text{eff}}$ is the effective number of lines producing the CC signal.

Equation \ref{eqn:hdc_ap} is written to highlight three terms.  The first compares the relative efficiencies of the two techniques (the denominator captures the loss of flux due to the lines).  The second captures the relative information content of the planet's spectrum in HDC. The third term is due to the speckle intensity and the lifetime of the speckles in AP.

A detailed parameter analysis is beyond the scope of the discussion.  Here we highlight two limiting cases.  The first is when $C_H$ is 10$\times$ larger than the limits achievable with optimized control of the atmosphere from \citet{males_guyon_2018}.  This large factor would be due to uncorrected quasi-static speckles, which have a large $\tau_{sl}$.  In this regime, the ratio is $>1$, indicating that HDC is the more efficient technique.  The result is shown in the top row of Figure \ref{fig:jared}, where we have used spectral type appropriate spectral templates.  We have also assumed a $3\lambda/D$ IWA.

The second case is when $C_H$ reaches the residual turbulence limit, and all long-lived quasi-static aberrations are suppressed.  In this case, AP is much more efficient than HDC.  We see the result in the bottom row of Figure \ref{fig:jared}, where we now assume a $1\lambda/D$ IWA. This clearly shows the potential of GSMTs, and motivates significant effort in breaking through the limits imposed by quasi-static speckles on current instruments.


\section{Priorities}\label{sec:priorities}
The scientific potential of direct imaging is extraordinary, and is in fact a key part of the science case that justified the GSMTs.  However, adequate resources must be deployed to advance the technology to reach our most ambitious goals.  The US community is concentrating much of its funding through NASA towards space-based direct imaging technology, but as we have shown in this white paper that given sufficient investment \textbf{\textit{there are significant opportunities for upgrading, developing, and deploying instruments on the ground that have the potential to reach these science goals}}. 

We believe that there is an excellent historical example of this kind of investment in the NSF Center for Adaptive Optics (CfAO), which significantly advanced the field of adaptive optics at a crucial period.  Interest is accumulating for a Science and Technology Center (STC) dedicated to advancing the state-of-the-art in high-contrast imaging for both direct imaging of exoplanets and other applications, such as microscopy for biological applications.  This STC, with potential international partnerships with Japan's Astrobiology Center (ABC) and ESO, would serve as a development hub for high-contrast imaging.  It could include funds to purchase telescope time and build common testbeds to enable technology development in the relevant environments.  Support from the NAS for technology investment for ground-based high contrast imaging could help enable such a Center.



\phantomsection

\bibliographystyle{apj}
\setlength{\bibsep}{0.0pt}
\scriptsize
\bibliography{literature}

\newcommand{\noopsort}[1]{}
\begin{thebibliography}{27}
\expandafter\ifx\csname natexlab\endcsname\relax\def\natexlab#1{#1}\fi

\bibitem[{{Bowler}(2016)}]{bowler_pasp_2016}
{Bowler}, B.~P. 2016, \pasp, 128, 102001

\bibitem[{Chew {et~al.}(2006)Chew, Clare, \&
  Lane}]{Chew_PySH_comparison_OptComm06}
Chew, T.~Y., Clare, R.~M., \& Lane, R.~G. 2006, Optics communications, 268, 189

\bibitem[{{Codona} \& {Kenworthy}(2013)}]{Codona-ShortExposure}
{Codona}, J.~L., \& {Kenworthy}, M. 2013, \apj, 767, 100

\bibitem[{{Delorme} {et~al.}(2016){Delorme}, {Galicher}, {Baudoz}, {Rousset},
  {Mazoyer}, \& {Dupuis}}]{Delorme-SelfCoherentCamera}
{Delorme}, J.~R., {Galicher}, R., {Baudoz}, P., {Rousset}, G., {Mazoyer}, J.,
  \& {Dupuis}, O. 2016, \aap, 588, A136

\bibitem[{{Dessenne} {et~al.}(1998){Dessenne}, {Madec}, \&
  {Rousset}}]{dessenne_etal98}
{Dessenne}, C., {Madec}, P.-Y., \& {Rousset}, G. 1998, \ao, 37, 4623

\bibitem[{{Dessenne} {et~al.}(1999){Dessenne}, {Madec}, \&
  {Rousset}}]{dessenne_etal99}
---. 1999, Optics Letters, 24, 339

\bibitem[{{Esposito} \& {Riccardi}(2001)}]{Esposito_Riccardi_AA2001}
{Esposito}, S., \& {Riccardi}, A. 2001, \aap, 369, L9

\bibitem[{{Frazin}(2013)}]{Frazin-ShortExposureSelfConsistent}
{Frazin}, R.~A. 2013, \apj, 767, 21

\bibitem[{{Fusco} {et~al.}(2015){Fusco}, {Sauvage}, {Mouilelt}, {Beuzit},
  {Dohlen}, {Petit}, {Beuzit}, {Suarez}, {Soenke}, {Downing}, {Baudoz},
  {Sevin}, {Baruffolo}, {Schmid}, {Salasnich}, {Puget}, {Feautrier}, {Rochat},
  {Moulin}, {Hugot}, {Vigan}, {Hubin}, \& {Puget}}]{fusco_etal15}
{Fusco}, T., {et~al.} 2015, in Adaptive Optics for Extremely Large Telescopes
  IV (AO4ELT4), E11

\bibitem[{{Guyon}(2005)}]{guyon05}
{Guyon}, O. 2005, \apj, 629, 592

\bibitem[{{Guyon} \& {Males}(2017)}]{guyon&males17}
{Guyon}, O., \& {Males}, J. 2017, ArXiv e-prints

\bibitem[{{Guyon} {et~al.}(2009){Guyon}, {Matsuo}, \& {Angel}}]{guyon_etal09}
{Guyon}, O., {Matsuo}, T., \& {Angel}, R. 2009, \apj, 693, 75

\bibitem[{{Johnson} {et~al.}(2011){Johnson}, {Gavel}, \&
  {Wiberg}}]{johnson_etal11}
{Johnson}, L.~C., {Gavel}, D.~T., \& {Wiberg}, D.~M. 2011, Journal of the
  Optical Society of America A, 28, 1566

\bibitem[{{Macintosh} {et~al.}(2005){Macintosh}, {Poyneer}, {Sivaramakrishnan},
  \& {Marois}}]{macintosh_etal05}
{Macintosh}, B., {Poyneer}, L., {Sivaramakrishnan}, A., \& {Marois}, C. 2005,
  in \procspie, Vol. 5903, Astronomical Adaptive Optics Systems and
  Applications II, ed. R.~K. {Tyson} \& M.~{Lloyd-Hart}, 170--177

\bibitem[{{Males} \& {Guyon}(2017)}]{males_guyon_2018}
{Males}, J.~R., \& {Guyon}, O. 2017, JATIS in press. ArXiv e-prints, 1712.07189

\bibitem[{{Martinache} {et~al.}(2014){Martinache}, {Guyon}, {Jovanovic},
  {Clergeon}, {Singh}, {Kudo}, {Currie}, {Thalmann}, {McElwain}, \&
  {Tamura}}]{martinache_etal14}
{Martinache}, F., {et~al.} 2014, \pasp, 126, 565

\bibitem[{{Mawet} {et~al.}(2017){Mawet}, {Ruane}, {Xuan}, {Echeverri},
  {Klimovich}, {Randolph}, {Fucik}, {Wallace}, {Wang}, {Vasisht}, {Dekany},
  {Mennesson}, {Choquet}, {Delorme}, \& {Serabyn}}]{mawet_HDCII_2017}
{Mawet}, D., {et~al.} 2017, ApJ, 838, 92

\bibitem[{{Perrin} {et~al.}(2003){Perrin}, {Sivaramakrishnan}, {Makidon},
  {Oppenheimer}, \& {Graham}}]{perrin_etal03}
{Perrin}, M.~D., {Sivaramakrishnan}, A., {Makidon}, R.~B., {Oppenheimer},
  B.~R., \& {Graham}, J.~R. 2003, \apj, 596, 702

\bibitem[{{Poyneer} {et~al.}(2007){Poyneer}, {Macintosh}, \&
  {V{\'e}ran}}]{poyneer_etal07}
{Poyneer}, L.~A., {Macintosh}, B.~A., \& {V{\'e}ran}, J.-P. 2007, Journal of
  the Optical Society of America A, 24, 2645

\bibitem[{Ragazzoni \& Farinato(1999)}]{Ragazzoni_sensitivity1999}
Ragazzoni, R., \& Farinato, J. 1999, Astronomy and Astrophysics, 350, L23

\bibitem[{{Riggs}(2016)}]{riggs16}
{Riggs}, A.~J. 2016, {KISS} Workshop: Exoplanet Imaging and Characterization:
  Coherent Differential Imaging and Signal Detection Statistics

\bibitem[{{Singh} {et~al.}(2014){Singh}, {Guyon}, {Baudoz}, {Jovanovich},
  {Martinache}, {Kudo}, {Serabyn}, \& {Kuhn}}]{singh_etal14}
{Singh}, G., {Guyon}, O., {Baudoz}, P., {Jovanovich}, N., {Martinache}, F.,
  {Kudo}, T., {Serabyn}, E., \& {Kuhn}, J.~G. 2014, in \procspie, Vol. 9148,
  Adaptive Optics Systems IV, 914848

\bibitem[{{Singh} {et~al.}(2015){Singh}, {Lozi}, {Guyon}, {Baudoz},
  {Jovanovic}, {Martinache}, {Kudo}, {Serabyn}, \& {Kuhn}}]{singh_etal15}
{Singh}, G., {et~al.} 2015, \pasp, 127, 857

\bibitem[{{Sivaramakrishnan} {et~al.}(2002){Sivaramakrishnan}, {Lloyd},
  {Hodge}, \& {Macintosh}}]{sivaramakrishnan_etal02}
{Sivaramakrishnan}, A., {Lloyd}, J.~P., {Hodge}, P.~E., \& {Macintosh}, B.~A.
  2002, \apjl, 581, L59

\bibitem[{{Snellen} {et~al.}(2015){Snellen}, {de Kok}, {Birkby}, {Brandl},
  {Brogi}, {Keller}, {Kenworthy}, {Schwarz}, \& {Stuik}}]{snellen_2015}
{Snellen}, I., {et~al.} 2015, A\&A, 576, A59

\bibitem[{Viotto {et~al.}(2013)Viotto, Magrin, Bergomi, Dima, Farinato,
  Marafatto, \& Ragazzoni}]{Viotto_PyramidBehavior2013}
Viotto, V., Magrin, D., Bergomi, M., Dima, M., Farinato, J., Marafatto, L., \&
  Ragazzoni, R. 2013, in Proc. Third AO4ELT Conf, Vol.~38

\bibitem[{{Wang} {et~al.}(2017){Wang}, {Mawet}, {Ruane}, {Hu}, \&
  {Benneke}}]{wang_HDCI_2017}
{Wang}, J., {Mawet}, D., {Ruane}, G., {Hu}, R., \& {Benneke}, B. 2017, AJ, 153,
  183

\end{thebibliography}

\end{multicols}
\end{document}